\documentclass{iaus}
\usepackage{graphicx}

\title[Disk Sizes in a $\Lambda$CDM Universe] 
{Disk Sizes in a $\Lambda$CDM Universe}

\author[Qi Guo \& Simon White]   
{Qi Guo$^1$
 \and Simon White$^2$}
\affiliation{Max Planck Institut for Astrophysics, \\ Postbus 85741,
 Garching, Germany \\ $^1$email: {\tt guoqi@mpa-garching.mpg.de} \\  $^2$email: {\tt	swhite@mpa-garching.mpg.de}}

\pubyear{2008}
\volume{xxx}  
\pagerange{?--?}
\jname{The Galaxy Disk in Cosmological Context}
\editors{A.C. Editor, B.D. Editor \& C.E. Editor, eds.}
\begin{document}

\maketitle

\begin{abstract}
We introduce a model which uses semi-analytic techniques to trace formation
and evolution of galaxy disks in their cosmological context. For the first
time we model the growth of gas and stellar disks separately. In contrast to
previous work we follow in detail the angular momentum accumulation history
through the gas cooling, merging and star formation processes. Our model
successfully reproduces the stellar mass--radius distribution and
gas-to-stellar disk size ratio distribution observed locally. We also
investigate the dependence of clustering on galaxy size and find qualitative
agreement with observation. There is still some discrepancy at small scale for
less massive galaxies, indicating that our treatment of satellite galaxies
needs to be improved.  
\end{abstract}

\firstsection 
\section{Introduction}

Understanding the origin of galaxy disks is an important aspect of the study
of galaxy formation and evolution. Almost all observed star formation occurs
either in galaxy disks or in material which came from galaxy disks (e.g. in
starbursts). The size of disks is closely related to their gas surface density
which in turn is critical in setting the star formation rate. Galaxy sizes
also correlate with many other physical properties, stellar mass, luminosity,
circular velocity, etc. These relations, as well as their evolution, pose
strong constraints on galaxy formation models.

A long-standing problem for cosmological simulations has been to reproduce
galaxies with the proper size distribution. The high efficiency of gas cooling
at early times, causes the later assembly of disks to proceed by coalescence
of cold gas clumps. As these clumps merge onto the main galaxy, they lose much
of their initial angular momentum through dynamical friction. The resulting
disks are then substantially smaller than observed and contain relatively
little stellar masss. Much effort has been directed to solving this problem,
most of it invoking some form of feedback to delay collapse
(\cite[Sommer-Larsen et al. 1999]{Sommer1999}, \cite[Thacker and Couchman
  2001]{Thacker2001}), or adopting an alternate initial fluctuation spectrum
with reduced small-scale power (\cite[Sommer-Larsen et al. 2001]{Sommer2001}).
Some authors (\cite[Governato et al. 2004]{Governato2004}) claim the problem
can at be significantly reduced by improving the numerical resolution, but
there is no consensus yet on the true solution.

In this work we implement a new treatment of disk formation within the Munich
galaxy formation model used in \cite[De lucia \& Blaizot 2007]{delucia2007},
hereafter DLB07. Our model distinguishes between gas and stellar disks and
treats their growth separately. We model the accumulation of mass and angular
momentum in a gas disk by conserving the specific angular momentum of the
cooling gas and the cold gas component of accreted satellite galaxies. When
star formation occurs in the disk, we further assume that the cold gas which
is converted into stars has the typical specific angular momentum of the
current gas disk. We apply our model to the Millennium Simulation and we
compare the stellar mass--size relation of disks and the distribution of the
ration of stellar and gas disk sizes to observation.  The dependence of galaxy
clustering on galaxy size is also investigated.

\section{Galaxy Formation Models}
\label{sec:models}
\subsection{Simulation}

The \emph{Millennium Simulation} is one of the largest simulations of cosmic
structure evolution so far carried out. It follows $N=2160^3$ particles within
a comoving box of side-length 685 Mpc from redshift $z=127$ to $z=0$. Each
particle has a mass of $8.6*10^{8}M_\odot$. The simulation assumes the
concordance $\Lambda$CDM cosmology with parameters consistent with a combined
analysis of the 2dFGRS (\cite[Colless et al. 2001]{Colless2001}) and the
first-year WMAP data (\cite[Spergel et al. 2003]{Spergel2003}): $\Omega_{\rm
  m}=0.25$, $\Omega_{\rm b}=0.045$, $h=0.73$, $\Omega_\Lambda=0.75$, $n=1$,
and $\sigma_8=0.9$, where the Hubble constant is parameterized as $H_0= 100h
\rm kms^{-1}Mpc^{-1}$. Galaxy formation models are then implemented on halo
merger trees constructed from the stored output of this dark matter
simulation. A detailed description can be found in \cite[Springel et
  al. 2005]{Nature2005}.
  
As the basis for our work, we use the galaxy formation model of DLB07,
changing only the treatment of galaxy sizes.  We refer the reader to the
original papers (\cite[Croton et al. 2006]{croton2006} and DLB07) for a
detailed description of this model.

\subsection {Disk size}

Here we introduce a new disk model by tracking the transfer of angular
momentum between the hot gas, the cold gas and the stellar component.

We assume that the hot gas cools onto the centre with specific angular
momentum which matches the current value for the dark matter halo. The change
in angular momentum of the gas disk can then be expressed as

\begin{equation}
\delta J_{g}=\dot{M}_{cool}\frac{J_{DM}}{M_{DM}}\delta t
\end{equation}
where $J_{g}$ is the total angular momentum of the gas disk, $\dot M_{cool}$
is the cooling rate, $\delta t$ is the corresponding time interval, and
$J_{DM}$ and $M_{DM}$ are the total angular momentum and total mass of the
dark matter halo, respectively.

In a minor merger, when the mass ratio of the two merging galaxies is larger
than 3, we assume that any cold gas in the satellite galaxy is added to the
disk of the central galaxy carrying a specific angular momentum equal to the
current value for the dark matter halo. The stars are added to the bulge. 
The angular momentum change is thus
\begin{equation}
\delta J_{g}={M}_{sat,gas}\frac{J_{DM}}{M_{DM}}
\end{equation}
where ${M}_{sat,gas}$ is the cold gas mass of the satellite galaxy. The final
angular momentum of the gas disk is a vector sum of its original angular
momentum, the changes due to gas cooling and accretion from
minor mergers, and the angular momentum lost to the stellar disk through star
formation (as described below).

For the stellar disk, we assume the star formation rate to be
proportional to gas density excess above some threshold. When cold gas is
converted into stars we assume it carries the average specific angular
momentum. 
\begin{equation}
\delta J_{*}=\dot{M}_{*}\frac{J_{g}}{M_{g}}\delta t=-\delta J_{g}
\end{equation} 
where $J_{*}$ is the total angular momentum of the stellar disk, $M_g$ is the
total mass of the gas disk, $\dot{M}_{*}$ is the star formation rate and
$\delta t$ is the corresponding time interval. As for the gas disk, the
angular momentum of the stellar disk is the vector sum of the original angular
momentum and the change due to star formation.

We assume the cold gas and stellar disks to be thin, rotationally supported
systems with exponential surface density profiles. For gas disk we have
\begin{equation}
\Sigma(R_g)=\Sigma_{g0}exp(-R_g/R_{gd})
\end{equation} 
and for stellar disk we have
\begin{equation}
\Sigma(R_*)=\Sigma_{*0}exp(-R_*/R_{*d})
\end{equation} 
where $R_{gd}$ and $R_{*d}$ are the scale-lengths and $\Sigma_{g0}$ and
$\Sigma_{*0}$ are the central surface densities for the gas and
stellar disks, respectively. Assuming the circular velocity to be constant, 
the galaxy to reside in an isothermal dark matter halo, and the gravity
of the galaxy to be negligible, the scale-lengths can be calculated as 
\begin{equation}
R_{g(*)d}=\frac{J_{g(*)}/M_{g(*)}}{2V_{cir}}
\end{equation} 
where $M_{g(*)}$ is the total mass of gas (stellar) disk and $V_{cir}$ is the
circular velocity. Here we
adopt the maximum velocity of the dark matter halo as $V_{cir}$ (\cite[Croton
  et al. 2006]{croton2006}).

\begin{figure}[h!]
\begin{center}
 \includegraphics[width=4.8in]{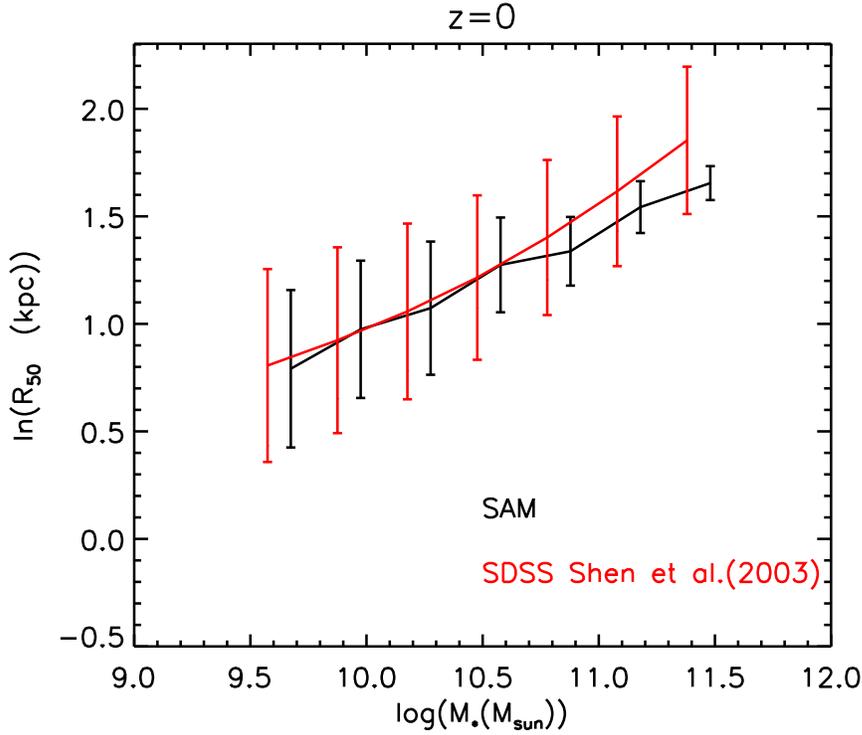} 
 \caption{Stellar mass vs. half stellar mass radius relation for spiral
   galaxies at the local universe. The black curve and black error
   bars are from our model catalogue and
   the red curve and red error bars are from SDSS data by \cite[Shen et al. 2003]{shen2003}}
   \label{size}
\end{center}
\end{figure}

\section{Results}
\label{effects}

In this section we show our predictions for the relation between stellar mass
and size, for the distribution of gas-to-stellar disk size ratios, and for the
dependence of galaxy clustering on galaxy size, and we compare them to
observation.
\subsection{Galaxy Stellar Mass and Size}
We select spiral galaxies from our model catalogue based on bulge-to-total
ratio. More specifically, we select all galaxies satisfying $1.5 \leq
\bigtriangleup M_I \leq 2.6$ ($\bigtriangleup M_I=
M_{Ibulge}-M_{Itotal}$). The relation between stellar mass and the
half stellar mass radius is shown in Fig.\,\ref{size}. Our model reproduces quite well the power-law dependence of galaxy
radius on stellar mass. Both the median value and the scatter match well for
galaxies less massive than $\sim 10^{10.5}M_{sun}$. For more massive galaxies,
the median value is still close to the observed value, but our scatter is much
smaller than observed.

\subsection{Gas-to-Stellar Size Ratio}

\begin{figure}[h!]
\begin{center}
 \includegraphics[width=4.8in]{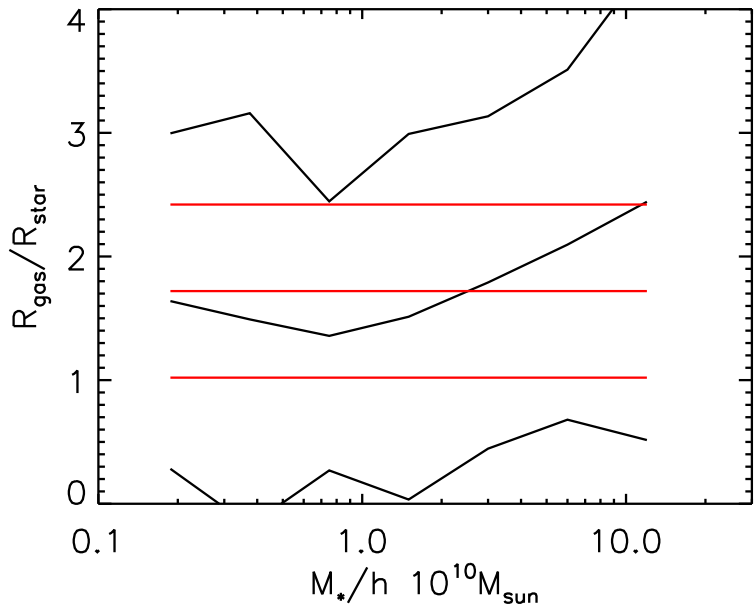} 
 \caption{The gas disk to stellar disk size ratio for Sa/Sab galaxies. As in
   Fig.\,\ref{size}, the black curves represent results from our model. The
   middle curve is the mean and the outer curves show one standard deviation
   from the mean.  Averaged over all Sa/Sab galaxies, the observational mean
   value is 1.72 (the middle red line) with scatter 0.70 (the other two red
   lines).}
   \label{ratio}
\end{center}
\end{figure}
Fig.\,\ref{ratio} shows the gas-to-stellar disk scale-length ratio as
   a function of stellar mass for Sa/Sab galaxies which are defined as
   $1 \leq \bigtriangleup M_I \leq 1.4$. There is almost no
dependence of the ratio on stellar mass. Averaging over all the galaxies, we
get a mean value of 1.67 and a standard deviation of 1.44, quite consistent
with observational results (the red lines in Fig.\,\ref{ratio}) reported in
\cite[Noordermeer at al. 2005]{Noor2005} based on the Westerbork HI survey of
spiral and irregular galaxies (WHISP).

\subsection{Correlation Functions}

\begin{figure}[h!]
\begin{center}
 \includegraphics[width=4.8in]{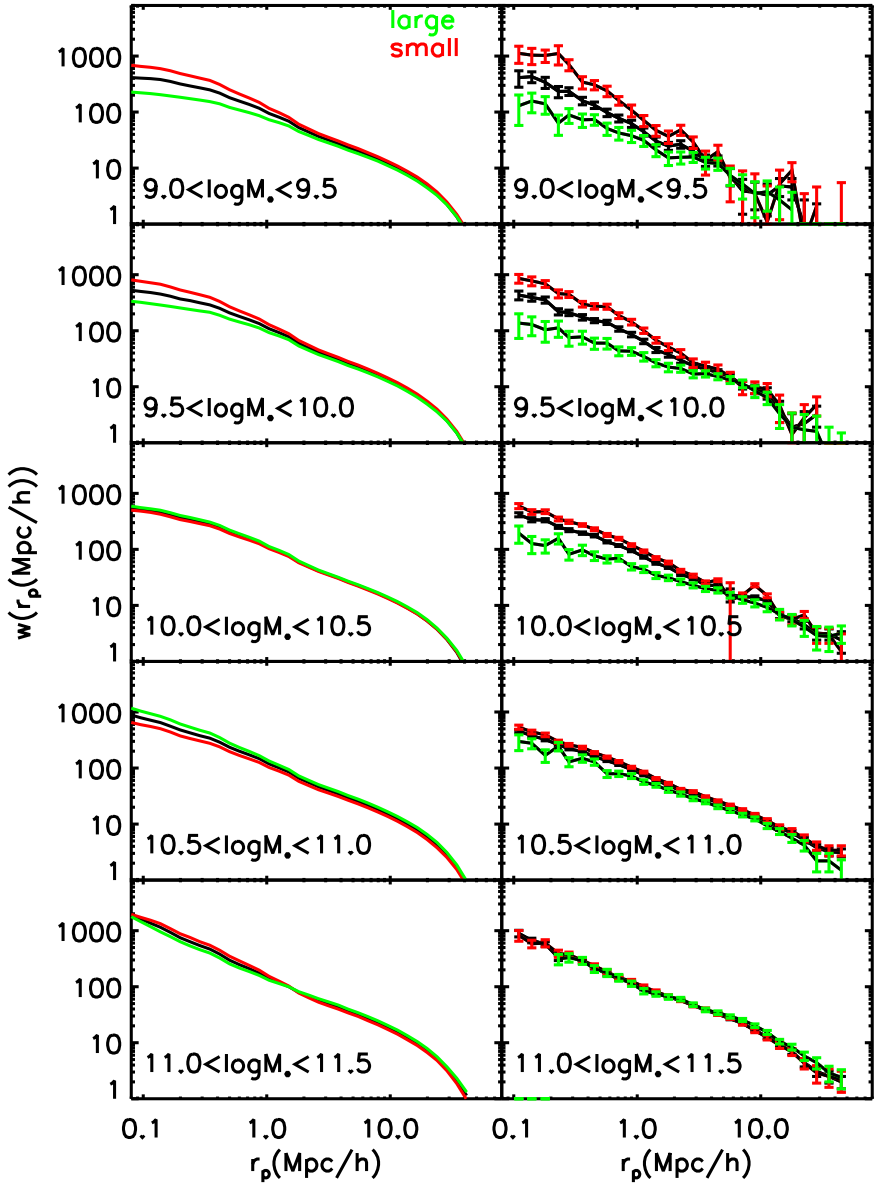} 
 \caption{Projected two point correlation function as a function of stellar
   mass and galaxy size. The left column is for our model and the right column
   is for SDSS data. The galaxy stellar mass bin is indicated at the bottom
   left corner of each panel. The projected correlation functions are
   represented by black curves.  In each mass bin the galaxies are further
   divided into large/low surface density (green) and small/high surface
   density (red) populations according to their half-mass radius/stellar
   surface mass density.}
   \label{cor}
\end{center}
\end{figure}
	
The dependence of clustering on galaxy size is shown in Fig.\,\ref{cor}. We
divide the galaxies into different mass bins and in each bin we further divide
the galaxies into two populations according to their sizes. On scales larger
than a few Mpc, there is no dependence of clustering on size for any stellar
mass. On smaller scales and for low-mass galaxies, the clustering of small
galaxies is much stronger than of large ones. The size dependence becomes
weaker with increasing stellar mass. For galaxies more massive than $\sim
10^{10}M_{sun}$, there is almost no dependence on galaxy size. 

These differences reflect size differences between central and satellite
galaxies of similar mass. For low-mass galaxies, centrals tend to be larger
than satellites of similar mass. This difference is much less for more massive
systems and probably reflects the facts that low-mass galaxies tend to survive
much longer as satellites than massive systems. The latter merge much more
quickly because of shorter dynamical friction times.  The size dependence of
our model galaxy is qualitatively consistent with the SDSS results taken from
\cite[Li et al. 2006]{li2006}. Quantitatively, however, the dependence is
stronger in the observational data and extends to more massive galaxies. A
possible way to remove this discrepancy would be to include the tidal
stripping of satellite galaxies in our model. This would reduce the sizes of
long-lived satellites further.

\section{Conclusion}

We have modeled the growth of gas and stellar disks in galaxies separately
using a semi-analytic approach which tracks the accumulation of their angular
momentum.  The differing angular momentum accumulation histories of galaxies
and of dark matter halos make it possible for a galaxy to have higher or lower
specific angular momentum than, and to be misaligned with, its surrounding
halo. The gas and stellar disks can also be misaligned, allowing the modelling
of galaxy warps.

We show that our model can reproduce many observed size-related relations
including the stellar mass vs. size relation and the gas-to-stellar disk size
ratio distribution. Both the mean values and the scatter are well reproduced,
especially for galaxies less massive than $10^{10.5}M_{sun}$. For the
dependence of clustering on galaxy size, we qualitatively reproduce observed
trends, but some discrepancy remains on the small scale for less massive
galaxies. Further improvement of our treatment of satellite galaxies,
including tidal stripping, may help to clarify the reasons behind this.

\end{document}